\documentstyle[12pt,epsfig]{article}
\newcommand{\be}{\begin{equation}}
\newcommand{\bea}{\begin{eqnarray}}
\newcommand{\eea}{\end{eqnarray}}
\newcommand{\ba}{\begin{array}}
\newcommand{\ea}{\end{array}}
\newcommand{\ee}{\end{equation}}

\expandafter\ifx\csname mathbbm\endcsname\relax

\else Kim

\fi
\def\appendix{{\newpage\section*{Appendix}}\let\appendix\section%
        {\setcounter{section}{0}
        \gdef\thesection{\Alph{section}}}\section}

\begin{document}

\begin{titlepage}
\hfill
\vbox{
    \halign{#\hfil         \cr
           CERN-TH/2000-386 \cr
           hep-th/0012222  \cr
           } 
      }  
\vspace*{30mm}
\begin{center}
{\Large {\bf  On Superconnections and  the 
Tachyon Effective Action}\\} 

\vspace*{15mm}
\vspace*{1mm}
Mohsen Alishahiha, Harald Ita and Yaron Oz 

\vspace*{1cm} 

{\it Theory Division, CERN \\
CH-1211, Geneva  23, Switzerland}\\

\vspace*{.5cm}
\end{center}

\begin{abstract}

We propose a form of the  effective action of
the tachyon and gauge fields for brane-antibrane systems and non-BPS Dp-branes, written
in terms of the supercurvature.
Kink and vortex solutions with constant infinite gauge field strength
reproduce the exact tensions  of the lower-dimensional
D-branes. 
We discuss the relation to BSFT and other models in the literature. 

\end{abstract}
\vskip 4cm

December 2000

\end{titlepage}

\newpage

\section{Introduction}
The open string tachyon condensation on non-BPS brane systems
 has attracted much interest recently.
One framework of analysis is level truncation of the open string field theory (SFT)
which lead to very good numerical agreements with expected values of vacuum energy and lower-dimensional
D-branes tensions \cite{lt,lt1}. Another framework is the boundary SFT (BSFT)
\cite{s,kutasov1,kutasov,a,lk,jap}.  It was argued that while in
the SFT approach an infinite number of  massive fields are 
involved in the condensation process, in the BSFT one can restrict to the tachyon field and study
some aspects of the condensation, such as the tensions of the lower-dimensional D-branes,
 exactly \cite{kutasov1,kutasov,k1}.

In this note we 
 will use the notion of superconnections \cite{quillen}, which
when 
considering the branes-antibranes system 
and  non-BPS Dp-branes
appears naturally via the Chan-Paton factors \cite{wittenk}. We will
make the assumption that the  effective action of
tachyon and gauge fields 
for the $Dp-\bar{Dp}$-branes system and non-BPS Dp-branes can be written
in a Quillen-like framework
in terms of the supercurvature.
The tachyon potential that arises in this framework is exponential in the tachyon field.
We will 
propose a form of the  effective action and use it to
study the process of tachyon
condensation. 
Kink solutions that we will find, with infinite constant value of the gauge field
strength,
reproduce the exact tensions  of the lower-dimensional
D-branes at the minimum of the tachyon potential. 
The effective action is different from the BSFT proposal \cite{kutasov,lk,jap}.
It can be related
by field redefinitions, in some cases, to the effective action proposed in 
\cite{GAsen,GA,GA1,GA2}.

The note is organized as follows.
In section 2 we will introduce the notion of superconnections and supercurvatures
and propose an effective action of
the tachyon and gauge fields for $Dp-\bar{Dp}$-branes system and non-BPS Dp-branes.
In section 3 we will study the 
kink solutions and derive
the exact tensions  of the lower-dimensional
Dp-branes. 
Section 4 is devoted to a discussion 
and comparison  the effective action to 
other models in the literature.

Note, that we will use a metric with signature $(-,+,...+)$.
Also, we will re-scale the gauge fields, tachyon and coordinates
by 
$A\rightarrow A/\sqrt{2\pi\alpha'},\;T\rightarrow T/\sqrt{2\pi\alpha'},\;x\rightarrow\sqrt{2\pi\alpha'}x$.
In section 3 we will re-scale back in order to get the correct dimensions and tensions.

\section{Superconnections and the effective action}
In this section we  will introduce  the notion of superconnections \cite{quillen}. We will
make the assumption that the  effective action of
tachyon and gauge fields 
for $Dp-\bar{Dp}$-branes system and non-BPS Dp-branes can be written
in a Quillen-like framework
in terms of the supercurvature, and 
propose the form of the  effective action.

\subsection{Supeconnections for $Dp-\bar{Dp}$ systems}

The superconnections which will be relevant for us
appear, for instance,  in the work of Quillen \cite{quillen}
on the Chern character of a K-class, and  
in the non-commutative formalism 
of Connes applied to to algebras of the form
${\cal C}^\infty({\cal R}^4) \otimes ( \,\,{I \kern -.6 em C} 
\oplus {I \kern -.6em C}\,\,)$ \cite{CL}.
We will mainly follow the formalism of Quillen, and briefly 
review in the following some of its elements.

One considers a pair of complex vector bundles $E_1,E_2$ over a manifold $M$ and a 
homomorphism $T:E_2 \rightarrow E_1$.
In the branes-antibranes system the vector bundles $E_1$ and $E_2$ correspond to the branes and antibranes
respectively, and the map $T$ corresponds to the tachyon arising from the open string stretched
between them. 
One can regard $E = E_1 \oplus E_2$ as a superbundle,
that is a bundle that carries
a $Z_2$-graded structure. The fiber $V = V_1 \oplus V_2$ is a vector space with $Z_2$ grading.
Denote
the involution that gives the grading by $\varepsilon$: $\varepsilon(v) = (-1)^{deg(v)}v$.  
The algebra of endomorphisms of $V$, $End(V)$, is a superalgebra with even and odd elements.
The even endomorphisms commute with  $\varepsilon$, while the odd ones anticommute with it.
The supertrace is defined by
\be
{\rm Tr}_s (X) \equiv Tr (\varepsilon X),~~~~X \in End(V) \ .
\label{supertrace}
\ee
It vanishes for odd endomorphisms, and gives the difference of the traces on $V_1$ and $V_2$
for the even ones.

When considering differential forms on $M$ there is a natural $Z$-grading corresponding to the degree
of the forms. Thus, differential forms on $M$ with values in $E$ have a $Z \times Z_2$ grading.
What will be relevant is the total $Z_2$ grading.

Let $D$ be an odd degree connection on $E$ preserving the $Z_2$ grading
\be
D =  \pmatrix{d+A^{1}  & 0 \cr &\cr
0 & d+A^{2}} \ .
\label{D}
\ee
Denote by ${\cal T}$ the odd degree endomorphism of $E$
\be
{\cal T}=  \pmatrix{0  & iT \cr &\cr
i{\bar T} & 0} \ .
\label{cT}
\ee
The supeconnection ${\cal A}= D + {\cal T}$ on $E$ is 
an operator of odd degree acting on differential form on $M$ with 
values in $E$
\be
{\cal A}=  \pmatrix{d+A^{1}  & iT \cr &\cr
i{\bar T} & d+A^{2}} \ .
\label{A}
\ee

When considering the branes-antibranes system the superconnection (\ref{A})
appears naturally via the Chan-Paton factors, where
the gauge fields of the branes $A_{\mu}^1$
and antibranes $A_{\mu}^2$ are the diagonal elements and the off-diagonal elements are the
tachyon $T$ and its conjugate $\bar{T}$. 
Note, that
while the   diagonal elements are 1-forms the off-diagonal
elements are 0-forms. However, the total grading of all the matrix elements in one.

The supercurvature ${\cal F}={\cal A}^2$ is given by 
\be
{\cal F}=  \pmatrix{F^1-T{\bar T}  & i{\cal D}T \cr &\cr
i\overline{{\cal D}T} & F^2-T{\bar T}} \ ,
\ee
where the covariant derivatives are defined by
\bea
&&{\cal D}T \equiv dx^{\mu} D_{\mu}T=dx^{\mu}(\partial_{\mu}T+A^{1}_{\mu}T-
TA^{2}_{\mu})\ , \cr
&&\cr
&&\overline{{\cal D}T} \equiv dx^{\mu}{\overline {D_{\mu}T}}=dx^{\mu}
(\partial_{\mu}{\bar T}+A^{2}_{\mu}{\bar T}-{\bar T}A^{1}_{\mu})\ .
\label{COV} 
\eea
$F^i, i=1,2$ are the gauge fields strength associated with the gauge potentials $A^i, i=1,2$. 
Note, that we used the fact that in this framework 
$T$ and ${\bar T}$ anti commute with $dx^{\mu}$.
The Chern character $ch(E_1) - ch(E_2)$ is represented by $Tr_s\; e^{{\cal F}}$ \cite{quillen}.

\subsection{$Dp-\bar{Dp}$ effective action}

One can rewrite the supercurvature (\ref{COV}) using the Clifford algebra. We replace
$dx^{\mu_1}...dx^{\mu_n}\rightarrow
\frac{1}{n!}\gamma^{\mu_1}...\gamma^{\mu_n}$, where
$\gamma^{\mu}$ satisfy the Clifford algebra 
$\{\gamma^{\mu},\gamma^{\nu}\}=2g^{\mu \nu}$.

The supercurvature reads now
\be
{\cal F}=\pmatrix{{1\over 2}\gamma^{\mu \nu}F^1_{\mu \nu}- 
(T{\bar T} -m {\bar m}) &\;\;\;i\gamma^{\mu}D_{\mu}T\cr &\;\;\;\cr
i\gamma^{\mu}{\overline {D_{\mu}T}} &\;\;\; {1\over 2}\gamma^{\mu \nu}
F^2_{\mu \nu}- (T{\bar T} -m {\bar m})} \ ,
\label{F}
\ee
where 
$\gamma^{\mu \nu}={1\over 2}[\gamma^{\mu},\gamma^{\nu}]$, namely $
 dx^{\mu} \wedge dx^{\nu}\rightarrow\gamma^{\mu \nu}$. Note that in
(\ref{F}) we used the freedom to add a constant part, represented by 
$m{\bar m}$.
In the  
non-commutative formalism 
with algebras
${\cal C}^\infty({\cal R}^4) \otimes ( \,\,{I \kern -.6 em C} 
\oplus {I \kern -.6em C}\,\,)$, $m,\bar{m}$ correspond to the ${I \kern -.6 em C} 
\oplus {I \kern -.6em C}$ part  (see also \cite{ROE}).

There are two natural trace operations we can take over the Clifford algebra.
We denote by ${\rm tr}$ the one simply taken over the Clifford algebra 
elements, e.g.
${\rm tr}(\gamma^{\mu}\gamma^{\nu}) = 2^{[(p+2)/2]}g^{\mu\nu}$ in a 
$(p+1)$-dimensional space.  
We denote 
by ${\rm atr}$ the antisymmetric trace over
the Clifford algebra, e.g.  ${\rm atr}(\gamma^{\mu}\gamma^{\nu}) =  
{\rm tr} \gamma^{\mu\nu}$,
which leads naturally to the wedge-product structure.
We denote by
${\rm Tr}$ the one taken
over the matrix structure of ${\cal F}$, and ${\rm Tr}_s$ is as in (\ref{supertrace}).

Since the superconnection appears naturally in the description of the branes-antibranes system 
it is natural to ask 
whether we can write the effective action in terms of the supercurvature.
The first hint is the $Dp-D{\bar p}$ effective action up 
to second order, as computed in perturbative string theory \cite{pesando}
\be
S_2 = T_p \int d^{p+1}x  \left({1\over 4}F^{1 \mu\nu}F^1_{\mu\nu}+{1\over 4}
F^{2 \mu \nu}F^2_{\mu \nu} - D^{\mu}T{\overline{D_{\mu}T}}
-( T{\bar T} -m {\bar m})^2\right) \ ,
\label{YM}
\ee
where by $T_p$ we denote the tension of a BPS Dp-brane.
This action can be written as
\begin{equation}
S_2  = - \frac{T_p}{2^{[(p+2)/2]}}  \int d^{p+1}x {\rm Tr}({\rm tr} 
{\cal F}^2) \ .
\label{AC2}
\end{equation}
One may suspect then that the higher order terms in the effective action,
in the  slowly varying fields approximation, where we neglect
terms like $\partial^k F$ and $\partial^l T, l > 1$, could
be of the form 
${\cal F}^{n}, n > 2$.
We will work in the slowly varying fields 
approximation in the following. 
We will see in the next section
that this approximation is sufficient for
the analysis of some exact properties of the tachyon condensation. 

The second hint comes from the form of the Wess-Zumino (WZ) term of the  branes-antibranes system.
It can be written as 

\begin{equation}
S_{WZ}=
{\tau}\int d^{p+1}x{\rm Tr_s}({\rm atr} (\Gamma \;{\cal C}\; 
e^{{\cal F}})) \ , 
\label{AC}
\end{equation}
 where $\tau$ is a normalisation constant,  $\tau =\frac{e^{-m{\bar m}}\;\mu_p}{2^{[(p+2)/2]}}$
and $\mu_p = g_s T_p$.
$\Gamma$ is given by (see appendix A)
\be
\Gamma=i^{[\frac{p-1}{2}]}\pmatrix{\tilde\gamma &0\cr 0&\tilde\gamma}, ~~~~\tilde\gamma=i^{[\frac{p-1}{2}]}\gamma^0...\gamma^{p} \ ,
\ee
and  
\be
{\cal C}=\sum {1\over n!}\gamma^{\mu_1,\cdots\mu_n}
C_{\mu_1,\cdots\mu_n}
\ee
where  $C_{\mu_1,\cdots\mu_n}$ is an 
$n$-form corresponding to the RR n-form field.

In the language of differential forms (\ref{AC}) reads 
\begin{equation}
S_{WZ}=
{\mu_p e^{-m{\bar m}}}\int {\cal C} \wedge {\rm Tr_s}(\;e^{{\cal F}}) \ .
\label{ACW}
\end{equation}
This 
WZ action was 
proposed in \cite{KW}, and is expected in view of the discussion in section 2.1
and the fact that
D-branes charge is measured by the K-theory class.

The supercurvature ${\cal F}$
can be decomposed as 
\bea
{\cal F}&=&\pmatrix{{1\over 2}\gamma^{\mu \nu}F^1_{\mu \nu}
&\;\;\;
i\gamma^{\mu}D_{\mu}T\cr &\;\;\;\cr
i\gamma^{\mu}{\overline {D_{\mu}T}} &\;\;\; {1\over 2}\gamma^{\mu \nu}
F^2_{\mu \nu}}- ( T{\bar T} -m {\bar m})\pmatrix{{ 1\kern - .4em 1} &\;\;0\cr &\;\;\cr
0&\;\; { 1\kern - .4em 1}}\cr
&&\cr
&=&{\bar {\cal F}}-( T{\bar T} -m {\bar m}){ 1\kern - .4em 1}  \ .
\label{C}
\eea

Using this form of the curvature the WZ action (\ref{ACW}) can be written as
\begin{eqnarray}
S_{WZ}={\mu_p} \int d^{p+1}x e^{-T{\bar T}}{\cal C} \wedge
{\rm Tr_s}\left(\sum_{n\leq p+1}\frac{{\bar {\cal F}}^n}{n!}
\right) \ .\label{WZ1}
\end{eqnarray}
The WZ action (\ref{WZ1}) suggests that 
the tachyon potential is 
\be
V(T,{\bar T}) \sim  e^{-T{\bar T}} \ .
\label{TACP}
\ee
This is in accord with the effective field theory \cite{MZ1},
string field theory \cite{kutasov1}
and $\sigma$-model computations \cite{t}.

We now turn to the non-topological part of the branes-antibranes action, which we will
denote by  DBI. We expect to get the same tachyon potential
(\ref{TACP}) in the DBI part.
We now make the assumption that
we can write it via the supercurvature. 
Since the superconnection and supercurvature appear as part
of the structure of the system via the Chan-Paton factors one may expect this to be the case.
However, it is also possible that only the topological part
of the branes-antibranes action can be written using the supercurvature.
This is related to the question whether the superbundle structure is indeed
a structure of the brane-antibrane system or only of its topological part.
We will continue with the assumption, bearing in mind that we do not have
a proof for it. 

The requirement of being able to write the DBI part using the supercurvature,
together with the requirement of getting the 
 same tachyon potential
(\ref{TACP}) in the DBI part, uniquely fixes the DBI action to
\begin{equation}
S_{DBI} = -{\tau_0}\int d^{p+1}x {\rm Tr}({\rm tr} 
e^{{\cal F}}) \ .
\label{AC1}
\end{equation}
$\tau_0$ is a normalisation constant given by 
${T_p\over 2^{[(p+1)/2]}} = \tau_0 e^{m{\bar m}}$.

The order ${\cal F}^2$ of (\ref{AC1}) is precisely (\ref{YM}).
Using the form of the curvature (\ref{C}) we have
\begin{equation}
S_{DBI}=-\frac{T_p}{2^{[(p+2)/2]}}\int d^{(p+1)}x \;e^{-T{\bar T}}\;{\rm Tr}
\left({\rm tr}\;  e^{{\bar {\cal F}}}\right)\ .
\label{ACP} 
\end{equation}

Thus, the proposed effective
action of the branes-antibranes system, written in terms of the
supercurvature (\ref{C},) is  $S = S_{DBI} + S_{WZ}$, with  $S_{DBI}$ 
given by (\ref{ACP}) and  $S_{WZ}$ by (\ref{WZ1}).

\subsection{Superconnections for non-BPS Dp-branes}

Consider now the non-BPS Dp-branes, i.e. odd $p$ in Type IIA and even $p$ in Type IIB string theories.
A non-BPS Dp brane is obtained by orbifolding the $Dp-\bar{D}p$ system by
the $(-1)^{F_L}$ operation \cite{sen}.
The action $(-1)^{F_L}$ on the Chan-Paton factors is realized by the matrix $\sigma_1$,
which leaves $I, \sigma_1$ invariant and projects out $\sigma_3,i
\sigma_2$.
In this way one finds the invariant gauge superconnection 
\be
{\cal A}=  \pmatrix{d+A  & iT \cr &\cr
iT & d+A} \ ,
\label{Anb}
\ee
which means 
setting 
$A=A_1=A_2$ and 
$\bar{T}=T$ in  (\ref{A}). 

A mathematical framework to discuss these superconnections is that of \cite{quillen},
with the appropriate Chern character forms associated with the 
D-branes charges. As we will discuss, they multiply the RR-fields in $S_{WZ}$ of the non-BPS Dp-brane.
Let us briefly review the formalism.

Let $C_1=C\oplus C\sigma$ denote the a superalgebra with 
$\sigma$ having odd degree and $\sigma^2=1$. Tensoring the 
vector bundle $E$ associated with the non-BPS Dp-brane with the superalgebra 
$C_1$,  one constructs a super vector bundle $E'$
\begin{equation}
  \label{eq:supervectorbundle}
  E'=E\otimes C_1 \ .
\end{equation}
The algebra of endomorphisms of $E'$, $End(E')$, is a superalgebra,
whose elements can be written 
as $a+b\sigma$ with $a,b$ in $End(E)$. The supertrace on 
$End(E')$ is given by
\begin{equation}
  \label{eq:supertraceodd}
  Tr_{\sigma}(a+b\sigma)=Tr(b) \ .
\end{equation}
The superconnection on $E'$ is given by
${\cal A}=(d+A)+i\sigma T$, 
where  $\sigma=\sigma_1$ is the Pauli matrix. It  
coincides with (\ref{Anb}). 
The 
supercurvature ${\cal F} = {\cal A}^2$ reads
\be
{\cal F}=  \pmatrix{F-TT  & i{\cal D}T \cr &\cr
i{\cal D}T & F-TT} \ .
\label{sp}
\ee
The
differential 
forms $Tr_\sigma({\cal F}^n)$ are closed and of odd degree.

\subsection{Non-BPS Dp-branes effective action}

One can rewrite the supercurvature (\ref{sp}) using Clifford algebra as
\bea
{\cal F}&=&\pmatrix{{1\over 2}\gamma^{\mu \nu}F_{\mu \nu}
&\;\;\;
i\gamma^{\mu}\partial_{\mu}T\cr &\;\;\;\cr
i\gamma^{\mu}\partial_{\mu}T &\;\;\; {1\over 2}\gamma^{\mu \nu}
F_{\mu \nu}}- ( T^2 -m^2)\pmatrix{{ 1\kern - .4em 1} &\;\;0\cr &\;\;\cr
0&\;\; { 1\kern - .4em 1}}\cr
&&\cr
&=&{\bar {\cal F}}-( T^2-m^2){ 1\kern - .4em 1}  \ ,
\label{Cnon}
\eea
where we added a constant part.

Consider the WZ part of the non-BPS Dp-brane action.
It has been analysed partially in \cite{wz}.
The above discussion of the D-branes charges associated with the
non-BPS Dp-brane imply that the WZ-term is 
\begin{equation}
S_{WZ}=\frac{-i \mu_p}{\sqrt{2}} \int d^{p+1}x e^{-T^2}{\cal C} \wedge
{\rm Tr_{\sigma}}\left(\sum_{n\leq p+1}\frac{{\bar {\cal F}}^n}{n!}
\right) \ .
\label{eq:non-bps-WZ}
\end{equation}
That is, the Chern characters of the superbundle 
(\ref{eq:supervectorbundle}) encode the Dp-brane charges of 
condensates on the non-BPS branes. 
The WZ action suggests that 
the tachyon potential is $V(T) \sim  e^{-T^2}$.
Following the same reasoning as for the $Dp-\bar{D}p$ system we propose a form
of the DBI part of the 
non-BPS Dp-brane action, written in terms of the supercurvature as
\begin{equation}
S_{DBI}=-\frac{T_p}{\sqrt{2}2^{[(p+2)/2]}}\int d^{(p+1)}x \;e^{-T^2}\;{\rm Tr}
\left({\rm tr}\;  e^{{\bar {\cal F}}}\right)\ ,
\label{ACPnon} 
\end{equation}
with ${\bar {\cal F}}$ given by (\ref{Cnon}). 
The factor $1/\sqrt{2}$ in (\ref{ACPnon}) is due to the 
fact
that the tension of a non-BPS Dp-brane is $\sqrt{2}T_p$, and the matrix structure of ${\cal F}$.

To summarize, the proposed effective
action of the non-BPS Dp-brane is  $S = S_{DBI} + S_{WZ}$, with  $S_{DBI}$ 
given by (\ref{ACPnon}) and  $S_{WZ}$ by (\ref{eq:non-bps-WZ}).

\section{Tachyon condensation}
In this section we will study the process of tachyon condensation in
the non-BPS Dp-brane and branes-antibranes systems.
We will derive the exact tensions for the lower-dimensional D-branes via kink and vortex
solutions,
where
the gauge field strength is an infinite constant. 

\subsection{Condensation on a non-BPS-brane}

Consider a non-BPS Dp-brane that carries a D(p-1)-brane charge.
Upon tachyon condensation we expect to get a BPS D(p-1)-brane.
The tachyon $T$ is a function of one coordinate transverse to the expected
D(p-1)-brane world volume. Denote this coordinate by 
$x_1=x$. We take the tachyon configuration to be $T = \alpha x$ where $\alpha$ is constant.
For such a configuration higher than two derivatives of the tachyon vanish and we do not have
to worry about not including them in the effective action.

Using the WZ action (\ref{eq:non-bps-WZ}) we get the 
coupling of RR $p$-form to the Non-BPS-brane. It reads 
\bea
S_{WZ}&=&\sqrt{2}\mu_p\int d^{p+1}x\; C_p\; \partial T\; e^{-T^2}\cr
&&\cr
&=&\mu_{p-1} \int d^px C_p \ ,
\eea
with $\mu_{p-1}=\mu_p\; 2\pi l_s$, and we have rescaled to restore the appropriate dimensions.
We see that we  get 
the charge corresponding to the D(p-1)-brane, independently of the form of the gauge field strength.

We assume a constant gauge field strength. With such a configuration
derivatives of the gauge field strength vanish and we do not have
to worry about not including them in the effective action.
For simplicity we will take only 
$F_{12}=F$ to be non-zero.

The supercurvature (\ref{Cnon}) reads
\be
{\bar {\cal F}}=\pmatrix{\gamma^1\gamma^2 F_{12} & i\gamma^1 \partial T\cr
&\cr i\gamma^1 \partial T & \gamma^1\gamma^2 F_{12} } \ .
\ee
We can evaluate the DBI action  (\ref{ACP}) exactly and get
\be
S_{DBI}=-\sqrt{2}T_p\int d^{p+1}x e^{-T^2}\cosh\sqrt{(\partial T)^2-F^2} \ ,
\label{special}
\ee
where $F=F_{12}$ and $F^2 \equiv  F_{12}F_{12}$.
The field equations read
\bea
&&\partial\left[e^{-T^2}\partial T\; \frac{\sinh(\sqrt{(\partial T)^2-F^2})}
{\sqrt{(\partial T)^2-F^2}}\right]+2Te^{-T^2}\cosh(\sqrt{(\partial T)^2-F^2})
=0 \ ,\cr
&&\cr
&&\partial\left[e^{-T^2}F\; \frac{\sinh(\sqrt{(\partial T)^2-F^2})}
{\sqrt{(\partial T)^2-F^2}}\right]=0 \ .
\label{feq}
\eea

When $\alpha$ is finite there is a solution of the field equations (\ref{feq})
with $F=0$. This solution does not reproduce the right tension of a D(p-1)-brane.
Indeed, we do expect to get the  D(p-1)-brane when $T$ is at the minimum of the potential.
This happens when $\alpha \rightarrow \infty$.
In this limit, a kink solution with finite non-zero tension requires $F$ to be infinite.
At 
the bottom of the tachyon potential the kinetic term of $F$ goes to zero. One may want to conclude
that it does not cost energy to change the value of $F$ and therefore it is not
of importance. However, this is correct only for finite changes of the value of $F$.
Infinite changes do cost energy and change the resulting tension of the
kink.   
There is a particular way of $F$ approaching infinity as $\alpha \rightarrow \infty$
that leads to a kink that satisfies  the BPS relation between
the charge and the tension.
The configuration profile is
\be
T=\alpha x, \;\;\;\;\; \alpha=\sinh(\sqrt{\alpha^2-F^2}) \ .
\label{con}
\ee
For this profile the equations of motion read
\bea
&&\alpha x
e^{-\alpha^2x^2}\left[\frac{\alpha^3}{{\rm arcsinh}\alpha}-\sqrt{1+\alpha^2}
\right]=0 \ , \cr
&&\cr
&&xe^{-\alpha^2x^2}\;\frac{\alpha^3\sqrt{\alpha^2-{\rm arcsinh}^{2}\alpha}}
{{\rm arcsinh}\alpha}=0 \ .
\eea
The field equations are solved  when $\alpha=0$ and 
$\alpha\rightarrow \infty$. The $\alpha=0$ solution means $T=F=0$
and corresponds to the top
of the tachyon potential where we have the non-BPS-brane.  The $\alpha\rightarrow \infty$
corresponds to the minimum of the tachyon potential.

Plugging the $\alpha\rightarrow \infty$ solution into the action (\ref{special}), we get
\be
S_{\rm kink}=-\sqrt{2}T_p\int d^{p+1}x e^{-\alpha^2 x^2}\sqrt{1+\alpha^2}|_{\alpha
\rightarrow \infty}=(-\sqrt{2}T_p)\sqrt{\pi}(\int d^py) \ .
\ee
Therefore the tension of the kink is $T_{\rm kink}=\sqrt{2\pi}\;T_p$. After
restoring the appropriate units
we have 
\be
T_{\rm kink}=(2\pi\sqrt{\alpha'})\;T_p  \equiv T_{p-1} \ ,
\ee
as the exact value.

As we noted, finite changes of the value of $F$ do not affect the tension of the kink, but
infinite changes will. All the other configurations will not satisfy the BPS relation between the
charge and the tension.

It is worth exploring the kink profile in another set of variables.
Consider the DBI action for the non-BPS Dp-brane proposed in 
\cite{GA}. It reads
\be
S=-T_p\int d^{p+1}x V(\tilde{T})\sqrt{-\det(\eta_{\mu\nu}+\tilde{F}_{\mu\nu}
+\partial_{\mu}\tilde{T} \partial_{\nu}\tilde{T})} \ .
\label{G}
\ee 
One can study the kink solutions via this action.
The special configuration (\ref{con}) corresponds to a kink solution of the
form $\tilde{T}=\alpha x, \tilde{F}=0$ in the variable of (\ref{G}), when
$\alpha \rightarrow \infty$.
Such a kink solution has been discussed in \cite{MZ1}. 
It is mapped to the variables of (\ref{con}) via
\be
{\tilde T}=T\ , \;\;\;\;(\partial {\tilde T})^2+{\tilde F}^2=
\sinh^2\sqrt{(\partial T)^2-F^2} \ ,
\label{map}
\ee
which maps, in our case, the action (\ref{G}) to (\ref{special}).

\subsection{Condensation on the brane-antibrane system}

Consider tachyon condensation on a $Dp$-$\bar{D}p$ system
carrying a D(p-2)-brane charge.
The tachyon should form a vortex-like configuration, with the topological charge of the vortex
 encoding the $D(p-2)$ brane charge. 

We take the tachyon configuration 
$T=\alpha z,\;\bar{T}=\bar{\alpha}\bar{z}$, where $z=x^1 + i x^2$.
 Inserting into the WZ action (\ref{WZ1}) we get the coupling of RR $p$-form to the BPS-brane. It reads
\bea
S_{WZ}^{(2)}&=&\mu_p\int d^{p+1}x \frac{1}{2\;p!}\epsilon^{\mu_0,...\mu_{p-1}\alpha\beta}C_{\mu_0...\mu_{p-1}}
\left( (F^{1}-F^2)_{\alpha\beta}+2D_{\alpha}T{\overline{D_{\beta}T}}
\right)e^{-T{\bar T}}\cr
&&\cr
&=&\mu_p\;(2\pi) (1+\Delta F)\int d^{p-1}x\frac{1}{p!}\epsilon^{\mu_0...\mu_{p-1}}C_{\mu_0...\mu_{p-1}} \ ,
\eea
where $\Delta F = F^1-F^2$.
Reinstalling $2\pi\alpha'$ one thus finds $\mu_{cond}=2\pi\;\mu_{p-2}(1+\Delta F)$.
Assume that only $F_{12}^i, i=1,2$ is different from zero. 
In order to find the 
exact charge  the vortex-like solution should have $F_{12}^1-F_{12}^2=0$.

In this setup the supercurvature (\ref{C}) reads 
\be
{\bar {\cal F}}=\pmatrix{\gamma^1\gamma^2 F_{12} & i(\gamma^1+i\gamma^2) \partial_{z} T\cr
&\cr i(\gamma^1+i\gamma^2) \partial_{\bar{z}}\bar{T}&\gamma^1\gamma^2 F_{12} }.
\ee
Evaluating from this the action (\ref{ACP}) we get 
\be
S=-2T_p\int d^{p+1}x e^{-|T|^2}\cosh\sqrt{2\partial_z T\partial_{\bar{z}}\bar{T}-F^2} \ .
\ee
The field equations read 
\bea
&&\partial_z\left[\partial_{\bar{z}}\bar{T}e^{-|T|^2}\frac{\sinh(2|\partial T|^2-F^2)^{1/2}}{(2|\partial T|^2-F^2)^{1/2}}\right]
-\bar{T}e^{-|T|^2}\cosh(\sqrt{2|\partial T|^2-F^2})=0,\cr
&&\cr
&&\partial_z\left[e^{-|T|^2}F_{z\bar{z}}\frac{\sinh(2|\partial T|^2-F^2)^{1/2}}{(2|\partial T|^2-F^2)^{1/2}}\right],
\eea
where we denote $ F=F_{12}=-\frac{i}{2}F_{z\bar{z}}$.

To calculate the vortex tension consider the following kink profile
\be
T=\alpha z, \;\;\;\;\; \beta=\sinh(\sqrt{2|\alpha|^2-F^2}) \ .
\ee
For this profile the equations of motion read
\bea
&&\alpha z
e^{-|\alpha|^2|z|^2}\left[\frac{|\alpha|^2\beta}{{\rm arcsinh}\beta}-\sqrt{1+\beta^2}
\right]=0\cr
&&\cr
&&|\alpha|^2ze^{-|\alpha|^2|z|^2}\;\frac{|\alpha|^2\beta \sqrt{|\alpha|^2-{\rm arcsinh}^{2}\beta}}{{\rm arcsinh}\beta}=0 \ .
\eea
Again, this profile can solve the equations of motion for $\alpha=0$ and $F=0$. This solution corresponds to the top
of the potential where we have the $Dp$-$\bar{D}p$ system. 
Condensation of the $Dp$-$\bar{D}p$ system to a $D(p-2)$ brane,
 corresponds to non zero fields with the tachyon mostly sitting at the minimum of 
the potential.
This happens for $|\alpha|\rightarrow\infty$.  $F$ has to be sent to infinity
such that the $D(p-2)$ tension saturates the BPS bound.

The correct scaling for the field strength $F$ can be found from calculating the tension of the vortex.
Plugging the $\alpha\rightarrow \infty$ solution into the action, we get
\be
S|_{\rm vortex}=-2T_p\int d^{p+1}x e^{-|\alpha|^2|z|^2}\sqrt{1+\beta^2}
=-2\pi T_p\;\frac{\sqrt{1+\beta^2}}{|\alpha|^2}\int d^{p-1}x \ .
\ee
Scaling F such that $|\beta|\rightarrow|\alpha|^2$ the tension of the vortex is $T_{p-2,cond}=2\pi\;T_{p-2}$. After reinstalling $2\pi\alpha'$ one finds 
\begin{equation}
T_{p-2}=(2\pi)^2\alpha'\;T_{p-2} \ ,  
\end{equation}
which is 
the correct value of the D(p-2)-brane tension.

\section{Discussion}

An obvious question is what is the relation between 
the effective actions proposed here  and others in the literature.
The first thing one may try is to relate them by fields redefinition.
While this can be done, at least in some cases, it is not clear
how meaningfull it is.
In the BSFT picture,  
an exact map means that we expect the same whole $2d$-flow from the UV fixed point corresponding
to the top of the tachyon potential to the IR fixed point corresponding
to the bottom  of the potential.
There is no reason why this should be the case.
For instance, in both actions (\ref{ACPnon}) and (\ref{G}) we neglect higher derivative terms of the tachyons
and the gauge fields and a precise map is likely to require these. 
It may be more natural to expect that they agree only at the fixed points, namely on-shell from
string theory viewpoint. 
One example of fields redefinition is  
the map (\ref{map}) which maps
the action (\ref{special}) to (\ref{G}). 
Indeed the kink solutions of both actions agree when $\alpha=0$
and  
$\alpha \rightarrow \infty$. 
However, the space of solutions of the field equations differ for finite values of the fields, 
as one can easliy
verify.

Setting the gauge fields to zero, $A^1=A^2=0$, we get
\be
S=-{2T_p\over 2^{[(p+2)/2]}}\int d^{(p+1)}x e^{-T{\bar T}}{\rm  tr}
\left(\cosh(\sqrt{\gamma^{\mu}
\gamma^{\nu}\partial_{\mu}T\partial_{\nu}{\bar T}})\right) \ .
\label{TACAC}
\ee
Consider, for instance, non-BPS Dp-brane case
where the
tachyon is real $T={\bar T}$.
Up to two derivatives the action has the structure familiar from BSFT
and $\sigma$-model
perturbation theory
\be
S=-2T_p\int d^{(p+1)}x e^{-T^2}(1 + \frac{1}{2}\partial_{\mu}T 
\partial^{\mu}T) \ .
\ee
However, it is easy to see that the numerical coefficients of the higher 
derivative terms do not match
those of the BSFT action \cite{kutasov}.
Indeed, in \cite{kutasov} one studies the tachyon condensation with only the tachyon field
excited and one gets the precise tension of the lower-dimensional D(p-1)-brane.
In our variables, we needed a nonzero configuration of the gauge field strength
in order to derive the precise tension of D(p-1)-brane from the kink solution.
Upon addition of the gauge fields in the BSFT formalism \cite{lk,jap} there is still a 
difference between the actions.

One can 
also set 
$T={\bar T}=A^2=0$ in the action (\ref{ACP}), which leads to
\be
S=-{T_p\over 2^{(p+1)/2}}\int d^{(p+1)}x\;{\rm tr}\;e^{{1\over 2}
\gamma^{\mu\nu}
F_{\mu\nu}} \ ,
\label{NONA}
\ee
which one can map to the DBI action 
\be
S=-T_p\int d^{p+1}x\sqrt{-\det(\eta_{\mu\nu}+{\tilde F}_{\mu\nu})} \ ,
\label{DBIA}
\ee
by relating $\tilde{F}$ to a formal expansion of $\sinh(F_{\mu\nu})$.
However, as above, one would expect a change of variables to include
the higher derivative terms which were neglected in the slowly varying fields approximation.

\vskip 1cm
{\bf Note added:} While typing the paper we received \cite{lk} and \cite{jap} which 
contain an overlap regarding the WZ part of the non-BPS Dp-brane action.

\newpage

\appendix{Clifford algebra conventions}

In this appendix, the special faithful representations of the Clifford algebra, we used in
 this note, will be constructed.

Let $C(p,q)$ denote the Clifford algebra with $(\gamma^{i})^2=-1,\; i=1,...,p$ and $(\gamma^{j})^2=1,\; j=p+1,...,q$. The representation for $C(1,p)$ with $p+1$ even is constructed as follows: Choose the Pauli matrices as a representation of $C(0,2)$
\begin{equation}
  \label{eq:paulimatrices}
  \sigma^1=\left(\begin{array}{ll}0&1\\1&0\end{array}\right),\quad\sigma^2=\left(\begin{array}{ll}0&i\\-i&0\end{array}\right),\quad\sigma^3=\left(\begin{array}{ll}1&0\\0&-1\end{array}\right).
\end{equation}
and an arbitrary faithful representation for $C(1,p-2)$,
\begin{eqnarray}
  \label{eq:representationeven}
  C(0,2)&=&\{\sigma^1,\sigma^2\},\\
  C(p-2,1)&=&\{\gamma^1,...,\gamma^{p-2},\gamma^0\},
\end{eqnarray}
then one finds 
\begin{eqnarray}
  C(1,p)&=&\{i\sigma^3\otimes \gamma^0,...,\;i\sigma^3\otimes \gamma^{p-2},\;\sigma^1\otimes{\bf1},\;\sigma^2\otimes{\bf1}\}.
\end{eqnarray}

The minimal faithful representation for odd dimensions can be written as the direct sum of the two inequivalent representations generated from the even dimensional representation in one lower dimension.
\begin{eqnarray}
  \label{eq:odddimensional}
  C(1,p+1)=\{\left(\begin{array}{ll}\gamma^0&0\\0&\gamma^0\end{array}\right),...,\;\left(\begin{array}{ll}\gamma^p&0\\0&\gamma^p\end{array}\right),\;\left(\begin{array}{ll}\tilde\gamma&0\\0&-\tilde\gamma\end{array}\right)\},
\end{eqnarray}
with $\gamma^a$ denoting the Clifford of one dimension lower and $\tilde\gamma=i^{[\frac{p-1}{2}]}\gamma^0...\gamma^{p}$ the generalised $\gamma_5$.

\end{document}